# A Dynamic Multi Agent based scheduling for flexible flow line manufacturing system accompanied by dynamic customer demand

Danial Roudi

Department of Mechanical Engineering, Eastern Mediterranean University, Famagusta, North Cyprus Via 10, Mersin, Turkey

Roudi.Danial@gmail.com

Ali Vatankhah Barenji

Department of Mechanical Engineering, Eastern Mediterranean University, Famagusta, North Cyprus Via 10, Mersin, Turkey

Ali.vatankhah@cc.emu.edu.tr

Reza Vatankhah Barenji

Department of Industrial Engineering, Hacettepe University, Beytepe Campus 06800, Ankara, Turkey

Reza.vatankhah@hacettepe.edu.tr

Majid Hashemipour

Department of Mechanical Engineering, Eastern Mediterranean University, Famagusta, North Cyprus Via 10, Mersin, Turkey

*Abstract*—**Dynamic rescheduling decision-making problem is an important issue in modern manufacturing system with the feature of combinational computation complexity. This paper introduces a multi-agent based approach using the detailed process, provided by Prometheus methodology, which used for the design of a simultaneous dynamic rescheduling decision making for flexible flow line manufacturing system that working under dynamic customer demand. The application has been completely modeled with the Prometheus Design Tool (PDT), which offers full support to Prometheus Methodology. The proposed dynamic scheduling decision making system is developed for Automated UPVC door and Windows Company and can be support both static and dynamic scheduling.**

## I. INTRODUCTION

An automated UPVC doors and windows fabrication is processing of cutting, welding and assembling UPVC doors and windows. This fabrication is composed of several component with deferent shape and size, although requires hundreds of operation with entrant flow although consists of some workstations and each contains one or even more machines. This companies follow the flow line manufacturing system, therefore scheduling problem of this system is called flow line manufacturing cell-scheduling problem (FMCSP) [1] , additionally dynamic costumer demand (time based constraints) is added to problems of these companies so it can be consider that scheduling problem of this companies is a real time system problems [2].

A manufacturing scheduling problem with time-based limitation can be considered as the scheduling problem of a real-time system [3]. The real-time systems are defined as those systems in which the correctness of the systems depends on both logical and time-based correctness [4]. The logical correctness refers to the satisfactions of resource capacity constraints and precedence limitation of operations [5]. The time-based correctness, namely timeliness, refers to the satisfactions of the time-based constraints such as interoperation time-based constraints and due dates. According to the Hyun Joong Yoon et al [1] real time system can be divided into two kinds of deadline, hard deadline and soft deadlines. The real time system with hard deadlines is the system in which time-based correctness is critical, whereas the one with soft deadlines is the system in which time-based correctness is important, but not critical. The scheduling techniques of real time systems are divided into static scheduling (offline scheduling) and dynamic scheduling.

Static scheduling techniques are applicable to real time systems in which jobs are periodic. They perform offline feasibility or schedule ability analyses. For instance hybrid harmony [6] search is used for solving the multi object FMCSP or A rate monotonic scheduling algorithm is the best known static priority scheduling method, in which higher priorities are assigned to the jobs with a shorter period and also NP hard method is used more to solving FMCSP in the static scheduling problem [7].

Dynamic scheduling techniques are advantageous in a system that uncertainty such as aperiodic jobs and machine failures can be taken into consideration [8]. Dynamic scheduling techniques are divided into dynamic planning-based approaches and dynamic best effort approaches. In dynamic planning-based approaches, schedule ability is checked at run time when a job arrives, and the job is accepted only if timeliness is guaranteed [9]. On the other hand, dynamic best effort approaches do not check schedule ability at all. Hence, dynamic planning-based approaches are adequate for the real-time systems with hard deadlines, whereas the dynamic best effort approaches are adequate for those with soft deadlines. There are quite few studies which consider dynamic scheduling of manufacturing system, especially dynamic scheduling of manufacturing system by considering dynamic customer demand [10]. In the previous work by authors addressed a multi agent decision making system for flexible manufacturing system [4,8, 11, 12]. The proposed architecture is design for handling machine breakdown and optimizing machine utility in the FMS focus on reconfiguration of control system [13]. However, dynamic scheduling and dynamic customer demand is not considered in their previous studies.

This paper present a multi agent based dynamic scheduling decision-making system for automated door and Windows Company (is flexible assemble line) with considering dynamic customer demand. Dynamic behaviors in this company such as diversification of production and reconfiguration are taken into consideration. The multi agent dynamic scheduling system is developed based on Prometheus methodology $^{TM}$. Prometheus methodology is a general-purpose design methodology for the development of software agent systems in which it is not tied to any specific model of agency in software platform [14]. The multi agent based dynamic scheduling system is completely modeled by the Prometheus Design Tool (PDT), which offers full support to Prometheus Methodology. The proposed scheduling method is designed mainly for the work cell with time-based constraints, although it is applicable of keeping the work cell free from time-based constraints.

The rest of this paper is organized as follows: introduce multi agent system in dynamic scheduling in section 2; in section 3 the definition of case study and illustration of existing lacks in FLMs is expressed; the design of proposed multi agent system given in section 4; in section 5 decision making mechanism which is proposed the rescheduling algorithm of the system for dynamic customer demand; section 6 consist of discussion and conclusion.

II.     Literature review

Multi agent system, a branch in artificial intelligence, provides a new way for solving distributed, dynamic problems. Agent technology has been widely accepted and developed in implementation of the dynamic scheduling and distributed control system [6,15, 16]. Multi agent-based software platforms are usually endowed with distributed intelligent functions, and are becoming a key technology in new manufacturing systems built in a distributed manner, such as intelligent manufacturing systems (IMSs). Many researchers have attempted to develop agent-based architectures to support manufacturing activities [15]. Since the late 1980s, a number of researchers have applied agent technology to perform production planning and control on the shop floor [17]. V Kaplanoğlu [18] proposed a real time dynamic scheduling system based on the agent-based system, which has the advantages of less sensitive to fluctuations in demand or available vehicles than more traditional transportation planning heuristics (Local Control, Serial Scheduling) and provides a lot of flexibility by solving local problems . Kai-Ying Chen et al. [19] Applying multi-agent technique in multi-section flexible manufacturing system. To set up the dynamic dispatching rules, a distributed agent based system is implemented, which is assist the agents to choose suitable dispatching rule in pertaining dispatching region and also the whole-hog dispatching of manufacturing system resolved through the agents cooperation. Leitao [20] surveyed the literature in manufacturing control systems using distributed AI techniques. Up to now there isn't practical implementation of this approach in the real factory cases due to lack of standard way and there is a need for conducting more scientific research in this field.

The main aim of this paper is proposing approach for dynamic scheduling of flow line manufacturing system. Our proposed approach not only works in the dynamic environment but also work in the static environment although proposed approach could contribute to real manufacturing systems. Agent based systems have provided an excellent opportunity for modeling and solving dynamic scheduling problems. Agent-based models consist of rule-based agents which are dynamically interacting. The agents within the systems, in which they interact with, can create real-world-like complexity. Based on that observation, the context of this study is to schedule the machine and material handling system (MHs) by means of emphasizing flexibility in a flow line manufacturing system through Multi-Agent System (MAS) approach. The proposed scheduling scheme is designed by means of Prometheus TM methodology.

### III. Design of the proposed multi-agent system

Agent-based modeling and design differs from the conventional systems design. In the present study, multi-agent based dynamic scheduling system is designed by using Prometheus TM methodology [21]. Prometheus methodology is a general-purpose design methodology for the development of software agent systems in which it is not tied to any specific model of agency of software platform. The Prometheus methodology defines a detailed process for specifying, designing, implementing and testing/debugging agent-oriented software systems. In addition to detailed processes (and many practical tips), it defines a range of artifacts that are produced along the way. The Prometheus methodology consist of four steps, the first three steps of this methodology is in design of any agent oriented software are same but the last step namely implementation step is different, In this study Jack will be selected as platform for implementing proposed MAS in the future work.

*A. Steps of Prometheus methodology are as follows:*

a) The system specification phase focuses on identifying the goals and basic functionalities of the system, along with inputs (percepts) and outputs (actions).
b) The architectural design phase uses the outputs from the previous phase to determine which agent types the system will contain and how they will interact.
c) The detailed design phase looks at the internals of each agent and how it will accomplish its tasks within the overall system.

### IV. Case study:

YBG (Yaran Bahar Golestan) is a small enterprise that produces make-to-order UPVC doors and windows by automated machines. YBG is situated in the north of IRAN and provides doors and windows to representatives in all over IRAN. The company is made up of two main departments: an administrative office, located in a downtown area with major communication and commercial infrastructures that expedite relations with partners and clients and the promotion of new products, and a production facility located a few kilometers away in the industrial area. The production process involves the production of the frames of the windows/doors and several assemblies' phases, in addition, test and quality control phases are performed.

The window components, such as fittings, profiles, and glasses, are ordered to partner companies that manufacture them according to the windows/doors designs. The windows frames are manufactured in the YBG plant from first substances (UPVC profiles). Nearly to fifteen models of doors and windows are manufactured: Tilt and turn windows, slide hung, top light, sliding folding, center hinge/pivot and etc. Three profile qualities are available: high quality with two different colors, medium with five different colors and economic one with two different colors. First substance or all finished windows components are stored in four different warehouses. UPVC profiles are standard lengths long enough to manufacture the largest frame size and undergo quality control according to a defined protocol. Finished UPVC components arrive with quality control already certified.

The problems that exist in current scheduling and control architecture which can be potentially improved by multi agent based dynamic decision making are as follows:

- The manufacturing system is schedule by static scheduling which is located on an administrative office thus all the decisions are issued by this unit.
- The stations (machines) have no autonomous scheduling unit for their operations
- The system lacks the real time scheduling and is not flexible in the case of dynamic customer demand.
- The scheduling of this system in the dynamic environment is hard NP problem.

Development of a multi agent based dynamic decision system to address these problems is justified as follows:

- When the dynamic customer demand accrues, the dynamic decision making system can be schedule the system in the dynamic manner.
- Development of multi agent based dynamic decision system can be fined optimal scheduling in the machine fail disturbance.
- The proposed system makes autonomous station level
- proposed MAS create real time communication in the system

A. System Specification Design:

System specification phase is first part of Prometheus methodology, System specification design phase consist of four sub phases namely: Analysis Overview, Scenario Overview, Goal Overview, System Role Overview. Specification of system goals is design in the Goal Overview diagram, resulting in a list of goals and sub goals, with associated descriptors. This phase responsible for identification of, system goals, development of set of scenarios that have adequate coverage of the goals, functionalities that are linked to one or more goals, negotiations among the types of agents and scenarios of the system are determined. Figure 2 shows Goal Overview diagram of the system.

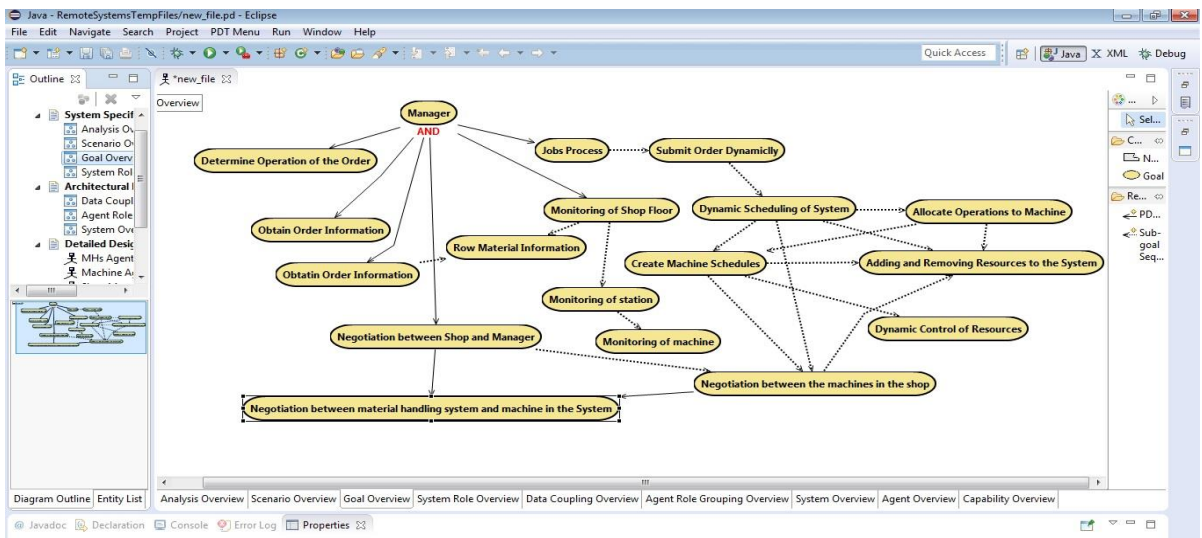

Figure 1 Shows Goal Overview Diagram of the System.

Scenario Overview was developed by set of scenarios that have adequate coverage of the goals, and which provide a process oriented view of the system to be developed. System Role Overview defined set of functionalities that are linked to one or more goals, and capture a limited piece of system behavior. Figure 3 shows System Role Overview of the System in which, there are four main roles in the system: Manger role, Shop Management role, cell role, Negotiation Management role.

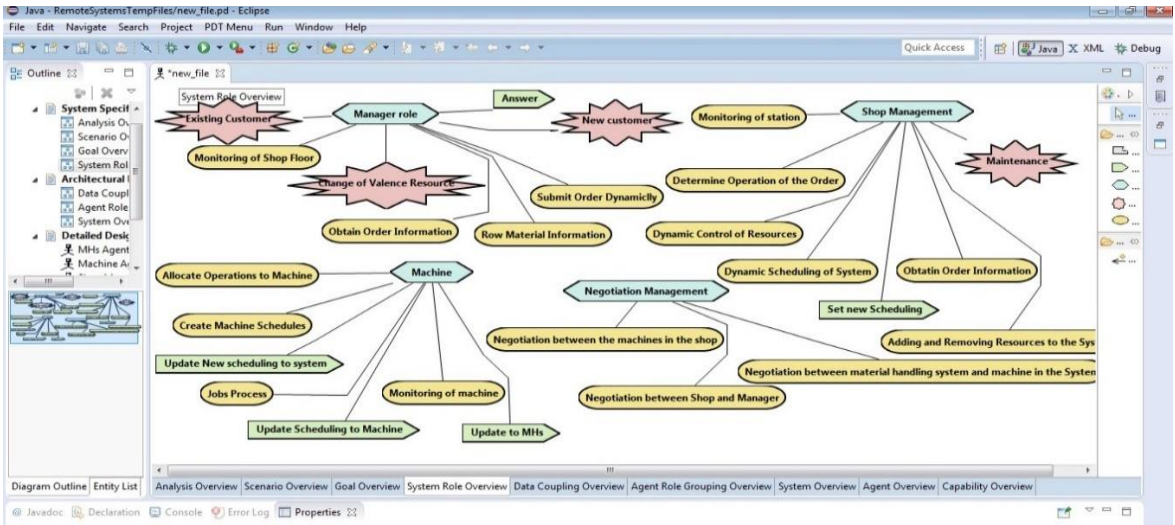

Figure 2 Shows System Role Overview of the System.

The sub goal is also designed in the system specification stage. For example four-sub goal of the Machine Scheduling after unpredictable orders arrived goal is defined; "Machine is busy and has task", "Machine is free and has a tack", "Machine is free and has no task", "Machine is loaded and has no tack".

B.  Architecture Design:

This stage includes the identification of agent types according to Prometheus methodology in which the roles of the agents in the system are determined. This phase consist of three parts namely; "Data Coupling Overview", "Agent Role Grouping Overview" and "System Overview". The negotiation protocols between agent types are designed in this phase. A system overview diagram is given in Figure 4. All agents are defined at this stage namely; "Manger Agent", "Shop Manager Agent", "Cell Agent", "MHs agent", "Scheduler Machine Agent", "MHs Resource Agent", "Machine Resource Agent". The last two agents are interface agents other five agents are software agents that used for dynamic scheduling decision-making system. The proposed system follow top to down approach but by considering real time negotiation between all types of agent. All negotiation protocols between agents are defined at Figure 4 diagram by arrows. Protocols consist of "Order Protocol", "Shop Protocol", "Material Handling System Negotiation Protocol", "Machine Negotiation Protocol", "Resource Protocol" and "Machine Resource Protocol".

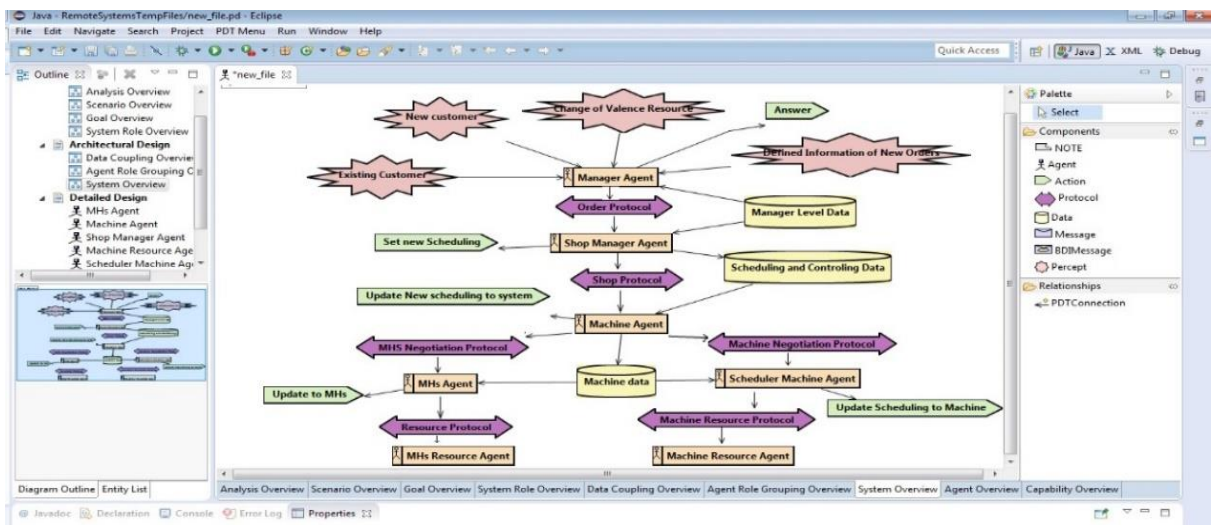

Figure 3 Shows System Overview Diagram in the Architectural Design Stage

*C.* Detailed Design:

Detailed design is done for each of the agent type one by one in this step. Types of agents at this stage take the message from the event of their environment or other agents, which operate on their plans, and thus they act according to the record in their data base.

For example Manager Agent (MA) is responsible to managing customer and updating new order to the system. Manager Agent uses its belief sets, plans and message events so that it will accomplish this task. Manager Agent architecture is shown in Figure 6 in the form of Prometheus ™ design view.

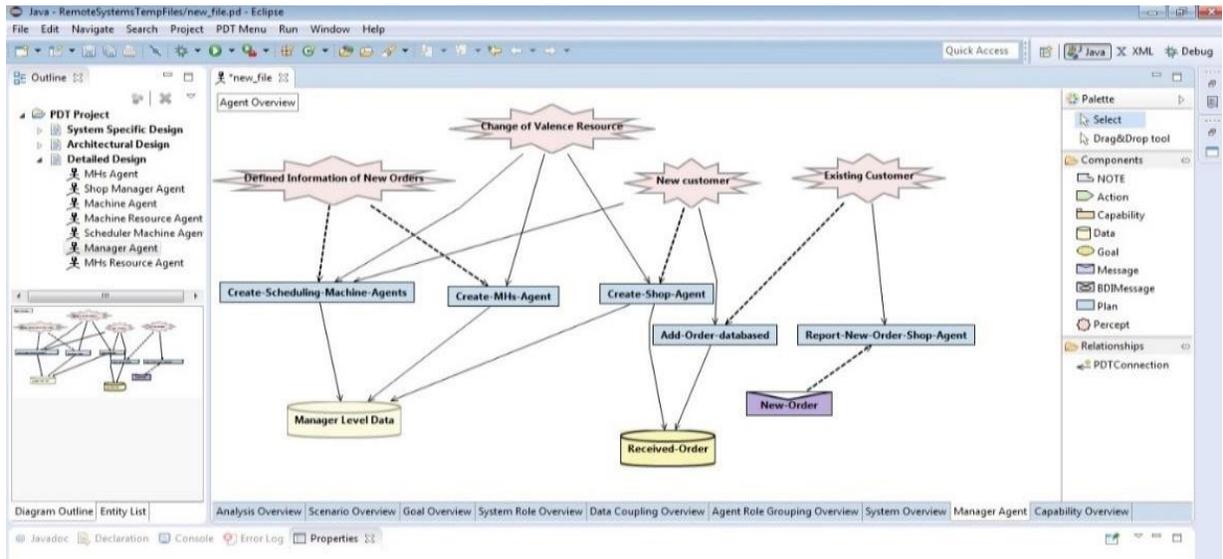

Figure 4 Manager agent architecture.

Machine agent is responsible for managing and controlling the cell level of factory and this agent consist of two sub agents namely MHs Agent and Scheduler Machine Agent. Detail design of this agent is in the Figure 7 having been showed.

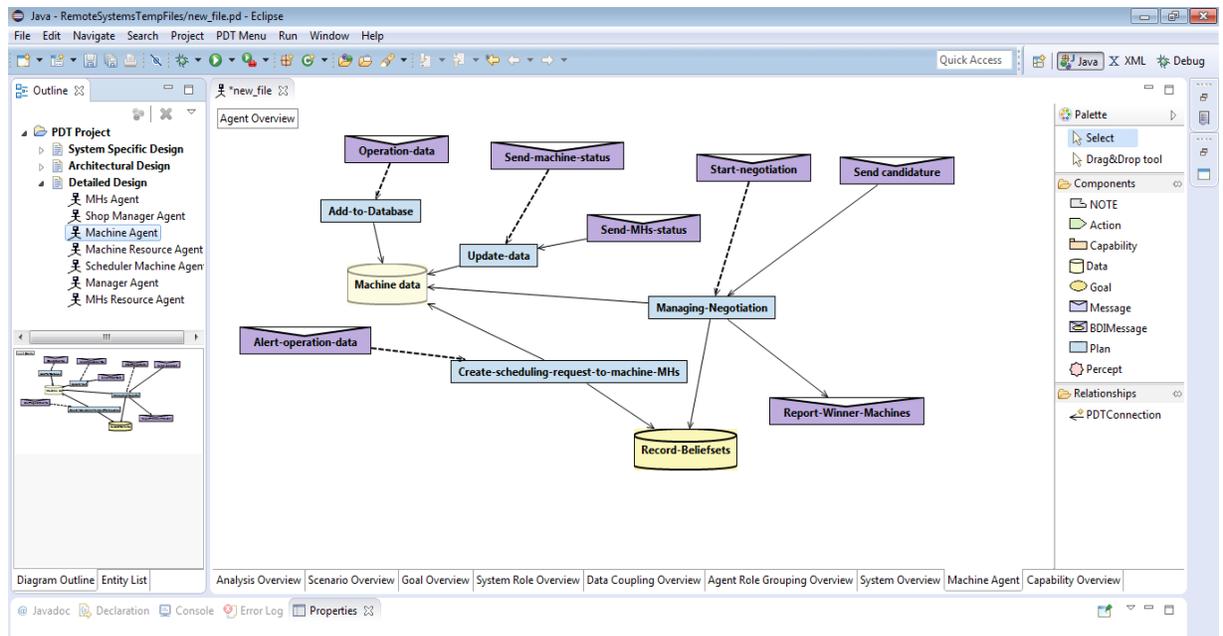

Figure 7 Detail Design of Machine Agent

Other important agent is play important role to rescheduling and dynamic scheduling of cell level is Scheduler Machine Agent. This agent consists of two data based Machine–status and Machine-Negotiation-Results. The detailed design of this agent illustrated in Figure 8.

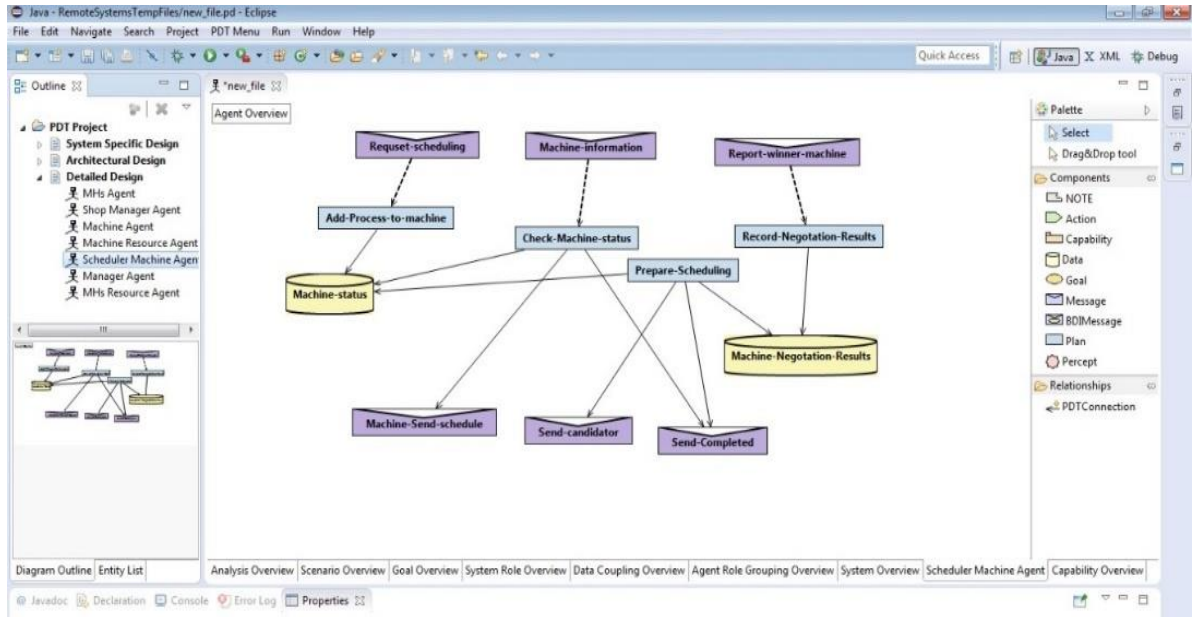

Figure 8 Detail Design of Scheduler Machine Agent

*D.* Decision making mechanism:

Algorithm for rescheduling of the system for dynamic customer demand is proposed at this section. Figure 9 is sequence diagram of decision-making mechanism in the proposed multi agent system. The Manager Agent informs the new or unpredictable order to the Shop Manager Agent. Shop Manager Agent send related question to the Cell Agent and this agent ask from Scheduler Machine Agent and MHs agent. Scheduler Machine Agent by having real time communication with Machine Resource Agent send related information to the Cell Agent and this agent by considering the information of Scheduler Machine Agent answers the question asked by Shop Manager Agent. Shop Manager Agent by considering information from cell level make a decision and send to Manager agent, if Manager Agent conform this decision, it will send related information to Shop Manger Agent, and Shop Manager Agent create new scheduling and new sub agent and send to the Cell Agent and MHs Agent. Cell Agent send new data to the scheduler Machine Agent and this agent create update new scheduler to the machine.

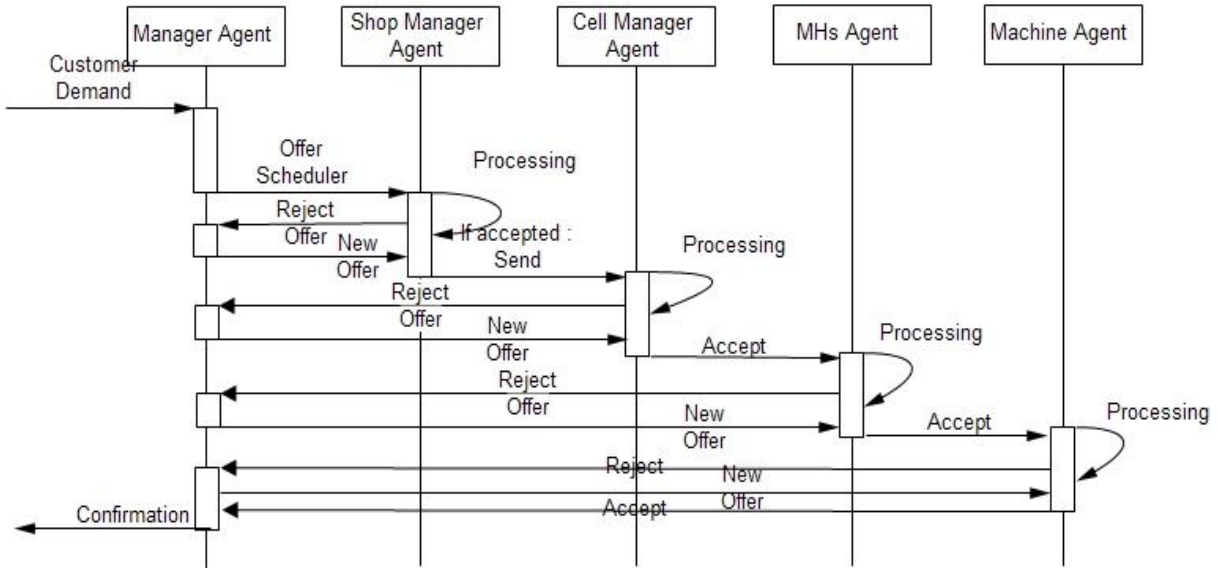

Figure 9 Sequence Diagram of Decision-Making Mechanism

V. Conclusion and future research

In the current study, the focus is on solving the scheduling problems of flow line manufacturing by proposing multi agent based decision-making system. The proposed multi-agent based design is developed in order to solve these complexities during the manufacturing process. The design uses the capabilities of multi-agent systems in order to solve real-time scheduling complexities. Feasible and effective schedules are supposed to be emerged from negotiation/bidding mechanisms between agents. In the present study, we try to make the problems of flow line manufacturing more clearly and how multi agent based system can be helpful for these types of companies. The other main of this study designing MAS based decision-making system for flow line manufacturing system by using Prometheus methodology.

Future research directions include

• Implementing the proposed MASs by using programming language.

• Finding test-bed studies in order to compare the results of multi-agent systems with other approximations.

• Developing simulation models in order to test the effectiveness of the proposed mode.

BIOGRAPHY

**Danial Roudi** is a Master of Engineering in Manufacturing Systems in the Department of Mechanical Engineering at the Eastern Mediterranean University, Famagusta, North Cyprus via 10, Mersin, Turkey. His research interests include manufacturing, simulation, optimization, reliability, and scheduling.

**Ali Vatankhah Barenji** has an MSc in mechanical engineering and is currently a PhD candidate at the University of Eastern Mediterranean. He is research assistance at the EMU. He has more than 15 papers published at international journal and proceedings of international conferences. His main research topics focus on reconfigurable production systems, intelligent supervisory control, and multi agent based control system.